\documentclass{emulateapj}
\usepackage{color}
\usepackage{courier}
\usepackage{sidecap}

\def\mgii{\ifmmode {\rm Mg{\sc ii}} \else Mg~{\sc ii}\fi}

\shortauthors{Smith, Mushotzky, Vogel, Shimizu \& Miller}
\shorttitle{Radio Properties of the BAT AGN}

\begin{document}

\title{Radio Properties of the BAT AGN: the FIR-Radio Relation, the Fundamental Plane, and the Main Sequence of Star Formation}

\author{Krista Lynne Smith\altaffilmark{1,2}, Richard F.~Mushotzky\altaffilmark{1}, Stuart Vogel\altaffilmark{1}, Thomas T. Shimizu\altaffilmark{1}, \& Neal~Miller\altaffilmark{3}}

\altaffiltext{1}{Department of Astronomy, University of Maryland College Park, MD 20742; klsmith@astro.umd.edu}

\altaffiltext{2}{NASA/GSFC, Greenbelt, MD 20771, USA}

\altaffiltext{3}{Department of Mathematics and Physics, Stevenson University, Stevenson, MD 21117}

\begin{abstract}

We have conducted 22~GHz 1\arcsec~JVLA imaging of 70 radio-quiet AGN from the \emph{Swift}-BAT survey. We find radio cores in all but three objects. The radio morphologies of the sample fall into three groups: compact and core-dominated, extended, and jet-like. We spatially decompose each image into core flux and extended flux, and compare the extended radio emission to that predicted from previous \emph{Herschel} observations using the canonical FIR-radio relation. After removing the AGN contribution to the FIR and radio flux densities, we find that the relation holds remarkably well despite the potentially different star formation physics in the circumnuclear environment. We also compare our core radio flux densities with predictions of coronal models and scale-invariant jet models for the origin of radio emission in radio-quiet AGN, and find general consistency with both models. However, we find that the $L_{\mathrm{R}} / L_{\mathrm{X}}$ relation does not distinguish between star formation and non-relativistic AGN-driven outflows as the origin of radio emission in radio-quiet AGN. Finally, we examine where objects with different radio morphologies fall in relation to the main sequence of star formation, and conclude that those AGN that fall below the main sequence, as X-ray selected AGN have been found to do, have core-dominated or jet-like 22~GHz morphologies.

\end{abstract}


\section{Introduction}
\label{sec:intro}

The apparent ubiquity of the correlation between the mass of a galaxy's central supermassive black hole and the stellar velocity dispersion of its bulge, known as the M$_{\mathrm{BH}}-\sigma$ relation \citep{gebhardt00, ferrarese00, gultekin09}, has prompted a robust investigation of the mechanisms by which these regions may influence one another. The concept of feedback in active galactic nuclei (AGN) attempts to explain the correlation by proposing a positive or negative regulatory effect of the AGN on star formation. In negative feedback scenarios, the AGN quenches the near-nuclear star formation by expelling or heating gas in the central regions \citep[e.g.,][]{dimatteo05, hopkins06, dubois13}. Positive feedback models suggest that AGN-driven jets or outflows generate turbulence and shocks which then trigger the collapse of giant molecular clouds and promote star formation \citep{klamer04}. The relationships may also be explained by co-evolution instead of ongoing interactions between the active nucleus and the nearby star-forming regions. Regardless of the physical method, it is now accepted that star formation and AGN activity are related phenomena. 

There is a nearly linear, remarkably tight relationship between the infrared and radio emission found in normal star forming galaxies, called the far-infrared (FIR)-radio correlation \citep{condon92}, which is widely used in studies of star formation out to high redshifts. It would be convenient to be able to apply this relationship to studies of star formation in AGN. This is complicated by the fact that even radio-quiet AGN tend to have nuclear radio emission in addition to extended emission from star formation. Further, a number of necessary assumptions are often employed in applying the FIR-radio correlation to AGN, including the use of star formation spectral energy distribution (SED) templates which assume that the AGN does not contribute to the far-IR emission. Recent work has shown that this assumption is flawed \citep{lira13}. It has also been shown that Seyferts have a wide variety of circumnuclear radio structures within the central kiloparsec, some attributable to star formation and some to linear structures resembling jets \citep[e.g.,][]{baum93}. It is even possible that the star formation environment surrounding the active nucleus may be quite different from that in isolated star forming regions in the host galaxy, and that circumnuclear star formation in AGN may not follow the FIR / radio relation at all. It is known that Seyferts tend to depart from this relationship, showing a radio excess that can be attributed to the AGN in the core \citep{wilson88,roy98,moric10}.  \citet{roy92} and \citet{baum93} showed that subtraction of the core AGN emission significantly improved the relation, but this was not borne out in the larger study by \citet{roy98}. 

In order to investigate this thoroughly, we have conducted a 22~GHz radio imaging program with the Karl G. Jansky Very Large Array (VLA) of 70 radio-quiet AGN from the ultra-hard X-ray \emph{Swift}-BAT survey \citep{baumgartner13}. With these high resolution (1\arcsec) radio images of a relatively unbiased AGN sample, we may spatially separate any emission due to star formation from the AGN core, and test the FIR-radio correlation using only the extended radio emission (beyond a few hundred parsecs). 

These observations also enable us to address another important mystery: the origin of the core radio emission in radio-quiet AGN. There seem to be very few, if any, radio-silent active galaxies. Radio observations with sufficient sensitivity have found compact radio emission in the majority of radio-quiet AGN \citep[e.g.,][]{nagar02,panessa10,maini16}. Ideas range from scaled-down versions of the powerful synchrotron jets seen in radio-loud AGN, to pure and highly compact star formation, to coronal synchrotron emission similar to that in active stars. Parsec-scale extended jet-like morphologies have been seen in some radio quiet AGN \citep[e.g.]{orienti10}; however, \citet{laor08} found that the relationship between the X-ray and radio luminosities of radio quiet AGN was consistent with an extension of the same relationship for coronally active stars, and \citet{baldi15} observed millimeter-band variability in the radio quiet AGN NGC~7469 consistent with X-ray variability, implying a common physical origin (presumably the corona). Correlations have been observed between the radio and X-ray luminosities in AGN of all types, and both of these quantities seem to be related to the supermassive black hole mass in a ``fundamental plane of black hole activity" \citep{merloni03}. The relationships between the X-ray luminosity, core radio luminosity and $M_{\mathrm{BH}}$ can place interesting constraints on accretion flow models and the geometry of the circumnuclear emitting regions. Using the same spatial decomposition that allows us to study the extended star formation in isolation, we see how the isolated AGN radio core properties of our sample fall on these relations.

The paper is organized as follows. In Section~\ref{sec:obs}, we discuss the properties of the sample and describe the detailed observations and reduction techniques. Section~\ref{sec:fluxes} discusses flux density measurements. Section~\ref{morph} focuses on the different star formation morphologies revealed by our high resolution maps. In Section~\ref{model}, we describe our methodology for decomposing the infrared SED into AGN and star formation components, and Section~\ref{condon} discusses our final results on the FIR-radio correlation. The fundamental plane and X-ray-radio correlation results are discussed in Section~\ref{origin}. Section~\ref{mainseq} examines where our samples of various radio morphologies fall on the main sequence of star formation. We discuss final conclusions in Section~\ref{sec:conclusion}.

Whenever redshifts and luminosities are discussed, we have assumed cosmological parameters $H_{0}=67.8~\mathrm{km s}^{-1}\mathrm{Mpc}^{-1}$ and $\Omega_{\mathrm{m}} = 0.308$, consistent with the most recent results from \emph{Planck} \citep{planck15}. 

\section{Observations and Reduction}
\label{sec:obs}
 
 \subsection{Sample}
Our parent sample is drawn from the 58-month version of the \emph{Swift}-BAT all-sky survey \citep{baumgartner13}. The survey was conducted in the ultra-hard X-ray band (14-195 keV). Because of its very high energy, this band is not affected by obscuration up to very high columns ($>10^{24}$~cm$^{-2}$). By the survey's 58-month catalog, it had detected $\sim600$ AGN of various types, many of which had never been detected as AGN at other wavebands. Hard X-ray selection is the least biased way to select AGN \citep{hickox09,koss11a}; the vast majority of all AGN are hard X-ray sources. Since optical, radio, infrared, and ultraviolet properties of the source do not enter into the selection, the sample is chosen independent of galaxy mass, galaxy luminosity, dust properties, radio loudness, or star formation rate. Additionally, most of the BAT AGN are relatively nearby, so high angular resolution translates to high spatial resolution for most of our sample. Our original work on the star formation in the BAT AGN was done in \citet{mushotzky14}, wherein we observed 313 of these objects with far-IR \emph{Herschel} PACS 160~$\mu$m and 70~$\mu$m images either from our Cycle 1 open-time program or the Herschel Science Archive. \citet{mushotzky14} attempted to separate the star formation and AGN contributions to the infrared SED using \emph{Herschel} imaging at 70 $\mu$m and 160 $\mu$m of a subset of the ultra-hard X-ray selected sample of \emph{Swift}-BAT AGN, but were unable to spatially distinguish the components with \emph{Herschel}'s 6\arcsec~resolution. It was clear that spatial decomposition would require higher resolution images. However, the AGN and the star formation can also be disentangled spectrally. By decomposing the \emph{Herschel} SED into AGN and star formation (SF) components, the methods in \citet{shimizu15} allow us to predict the radio emission expected from the FIR-radio correlation using the SF component of the FIR SED \emph{only}, putting us in a position to test the FIR-radio correlation if we obtained radio images with sufficient resolution to spatially decompose the AGN core and extended star formation. 
 
Many of the BAT AGN are detected in the FIRST \citep{becker95} and NVSS \citep{condon98} surveys at 1.4 GHz; however, these surveys have $\sim$5\arcsec and 45\arcsec~resolution, respectively. As indicated by the unresolved \emph{Herschel} images at 6\arcsec, higher resolution is needed to achieve the separation of the AGN and extended components. To this end, we initially obtained K-band (22~GHz) continuum observations of 45 of the \emph{Swift}-BAT AGN with the VLA in C-array, which has an angular resolution of $\sim$1\arcsec. We chose our initial sample to be unresolved in the \emph{Herschel} $70\mu$m images, and to be radio-quiet. All of our targets have $L_{1.4~\rm{GHz}} \leq 10^{23}~\rm{W}~\rm{Hz}^{-1}$, based on archival NVSS data. Additionally, we obtained VLA B-array follow-up observations of 17 objects which were unresolved in C-array, or which had a significant unresolved core and were sufficiently bright that B-array observations were feasible regardless of source structure. These images have a resolution of $\sim$0.3\arcsec. Although the B-array provides higher angular resolution, it also requires longer integration times to reveal extended structure, limiting the sample size for which we could obtain images of sufficient sensitivity. As described in Section~\ref{extended}, the majority of our B-array follow-up observations did not alter the morphological classification obtained with our C-array imaging. Finally, we obtained C-array observations in the following cycle for 25 new objects selected to be at low redshifts, to provide maximal spatial resolution and to supplement the original sample. Unlike the first set, this sample had no preselection to be unresolved in the \emph{Herschel} images. The final total observed sample is 70 targets. 

\subsection{Observations and Reductions}

All observations were conducted in the K-band, which has a central frequency of 22~GHz and a large 8~GHz bandwidth. We selected K-band (22~GHz) for our initial study rather than the more traditional lower frequency bands because we were aiming for 1\arcsec~angular resolution to spatially resolve the star formation and AGN emission, which was only possible at K-band in the available arrays (C and D arrays). We remained at K-band for our subsequent two sets of observations to provide a data set at a uniform frequency. The initial sample of 45 objects was observed in C-array in May 2013, the B-array sample was observed in September 2014, and the final low-redshift sample was observed in February 2015. Each object was observed in a one-hour block with 2-3 other objects. Each observing block began with attenuation scans, followed by flux and bandpass calibration on either 3C~48, 3C~138, or 3C~286, depending on sky position. Each object was observed for 3-8 minutes based on scheduling block constraints, and was followed before and after by a gain calibration scan. 

The data were processed using the Common Astronomy Software Applications package (CASA) \citep{mcmullin07}, which was developed to process interferometric and single-dish data from radio astronomical telescopes and is hosted primarily by the National Radio Astronomy Observatory (NRAO)\footnote{For more information on CASA, see www.casa.nrao.edu.}. 

After passing the raw data through the standard VLA reduction pipeline, we split off each individual object from the main measurement set, averaging over all 64 channels within each spectral window, resulting in a single data point per window. This reduces the amount of processing time without compromising the data. Then, each object was imaged with Briggs weighting (robust=0.5) and cleaned to a 0.03 mJy threshold or the dynamic range limit, whichever was higher. The images were then assessed for systematic errors; some exhibited significant radio frequency interference (RFI) signatures, which we removed by flagging the affected spectral windows. Images with a peak flux density exceeding 1~mJy went through iterative rounds of phase-only (non-amplitude) self-calibration of their visibility data (sources with lower peak flux densities did not have high enough signal-to-noise in the self-calibration solutions). Although the fractional bandwidth (~36\%) of the observations is quite large, standard multi-frequency synthesis using the CASA task CLEAN worked well because the angular extent of emission is less than 10 to 20\% of the primary beam width and because the achievable dynamic range of the data was modest (typically 100/1 or less). In most cases this is because our sources are fainter than 3~mJy (100 times the thermal noise, which is approximately 30$\mu$Jy). For the few sources with peak flux densities higher than 3~mJy, the dynamic range was limited by factors such as gain calibration errors that we were unable to correct. In cases where the S/N warranted it, we specified two terms in the multi-frequency imaging, providing spectral index maps.  In general the spatial spectral index variations were small and had no effect on the imaging. In the end, five images had to be discarded due to persistent RFI banding, and three objects were not detected at our sensitivity threshold (Mrk~653, Mrk~595, and 2MASX~J0107-1139). The final sample of useful images consisted of 62 objects, and is tabulated in Table~\ref{t:tab1}.

Lastly, we applied a Gaussian 6\arcsec~taper to the visibility data of each observation, to create a second image with resolution mimicking that of the original \emph{Herschel} observations. In the end, we have 62 images at 1\arcsec~resolution and 62 at 6\arcsec~resolution, which we later use to compare extended and core emission (see Section~\ref{condon}). 

\section{Flux Measurements}
\label{sec:fluxes}

 To measure the 22~GHz flux densities of unresolved or compact images and the compact cores of extended images, we used the CASA command \texttt{imfit} to fit an elliptical gaussian to our Stokes I image component. Although this method is questionable for use on extended sources, our compact sources are fitted very well by a single elliptical component. Each fit was inspected visually and manually re-done in the case that the elliptical was incorrectly fitted over an inappropriately large area to include larger-scale flux. 
 
A major objective in the project was to measure the extended emission in each image, which could be either star formation or an AGN-related jet or outflow. Although many of our 1\arcsec~resolution images exhibit large-scale structures even at very low surface brightness, we can be more confident of measuring this extended emission accurately by using our lower-resolution 6\arcsec~images. The extended flux in the higher resolution images is included in the 6\arcsec~beam in most cases. We can then use \texttt{imfit} on these unresolved large-scale images, and then subtract the core flux density as measured above in the full-resolution images, to calculate the value of extended flux only. In the handful of cases with extended flux visible in the 6\arcsec~images, it was measured manually in the CASA task \texttt{viewer}, using \texttt{imstat} on a box region drawn around the extent of the emission.

Of course, the emission inside the 1\arcsec~radio cores need not be exclusively caused by the AGN. For the full redshift range of our sample, 1\arcsec~resolution corresponds to 70 parsecs - 1 kiloparsec (see the following section). In order to be as conservative as possible, we measure the integrated flux density in the compact component of the image, rather than exclusively inside the unresolved beam, as the ``core flux density." (Note that the compact component is sometimes larger than the beam, and so is resolved, despite being compact in morphology). This core flux density may still include unresolved star formation and jets, as well as the nuclear AGN component. By measuring the integrated flux density in the compact component to be subtracted off, we ensure our estimates of the extended star formation flux densities are conservative lower limits; including the core emission would provide an upper limit. 

\section{Nuclear Star Formation Morphologies of \rm{BAT} AGN}
\label{morph}
The sample has a wide variety of 22~GHz morphologies. The resolution of the VLA at 22~GHz in C-configuration is $\sim$1\arcsec. Our objects are relatively nearby with a redshift range of $0.003\leq z\leq{0.049}$. This corresponds to a spatial scale of 0.07~kpc to 1~kpc. The redshifts of our target objects are shown in Table~\ref{t:tab1}. Of the 62 total objects, 34 are compact or unresolved, 22 show significant extension indicative of star formation or large outflows, and 6 are jet-like. We do not find any tendency for extended sources to be seen at lower redshifts, so this distribution is not a function of better resolution in nearby objects.
\\

\subsection{Core-Dominated Sources}
\label{compact}
 
 All of our detected galaxies have an unresolved or nearly unresolved core component. Additionally, all of our galaxies have emission that is extended relative to the 1\arcsec~beam. However, in approximately half (34 out of 62) of our sources the extended emission is too weak at 1\arcsec~to be significantly detected in individual resolution elements, and the detection of extended structure only becomes significant with smoothing or integrating. We refer to the sources in which extended emission requires smoothing for significant detection as ``core-dominated." 

The origin of this core emission is discussed in more detail in Section~\ref{condon} and Section~\ref{origin}. It is likely that many of our cores contain both AGN emission and unresolved star formation.


\begin{figure*}
    \centering
    \includegraphics[width=0.75\textwidth]{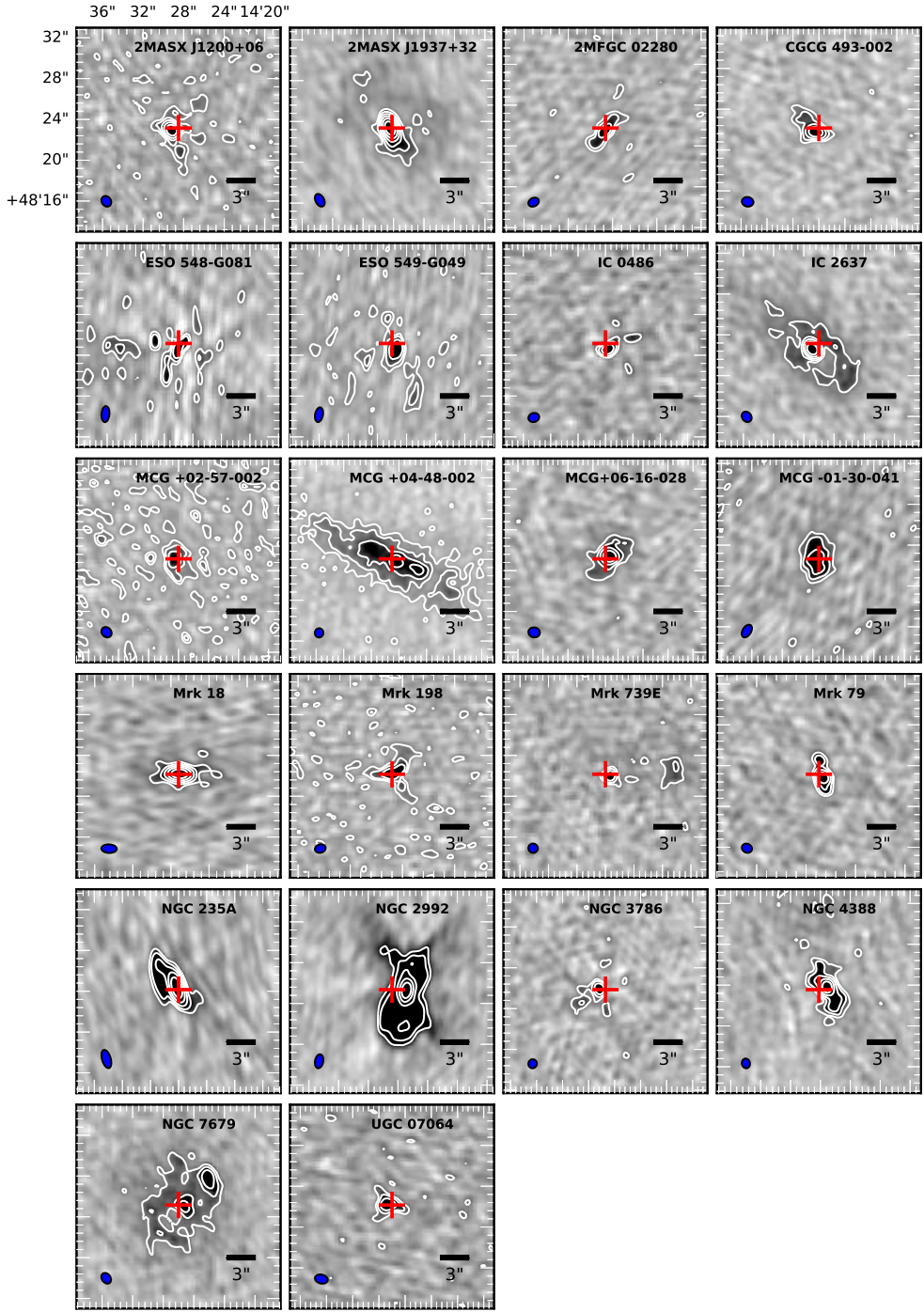}
    \caption{Radio continuum images at 22~GHz of the 22 radio-quiet BAT AGN with well-resolved, non-jetlike extended morphologies. The FWHM of the beam for each observation is shown in blue in the lower left corner of each image. Contours begin at 50\% of the peak flux density, and decrease outwards by factors of two. A red cross is shown at the phase center of each image, as well as a 3\arcsec~scalebar. Major ticks correspond to 4\arcsec, as can be seen by the scale given in the first panel.}
    \label{fig:extended}
\end{figure*}


\subsection{Star Formation, Outflows and Jets}
\label{extended}

\begin{figure*}
    \centering
    \includegraphics[width=0.7\textwidth]{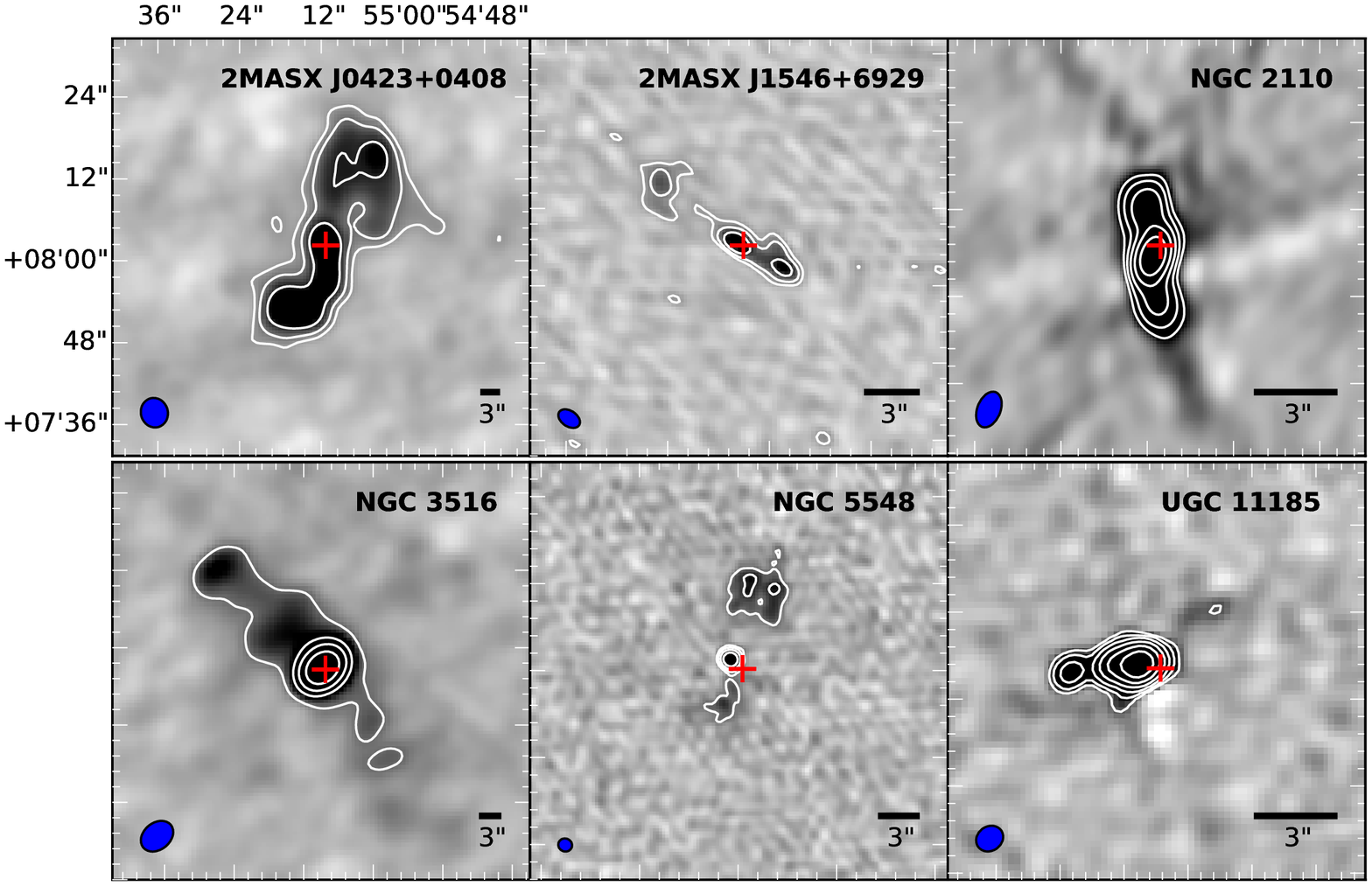}
    \caption{Radio continuum images at 22~GHz of the 6 radio-quiet BAT AGN with well-resolved, jetlike morphologies. The synthesized beam for each observation is shown in blue in the lower left corner of each image. Contours begin at 50\% of the peak flux density, and decrease outwards by factors of two. 2MASX~J0423+0408 and NGC~3516 are shown in 3\arcsec-tapered images to improve visibility of the jetlike structure. A 3\arcsec~scalebar is provided in each image, as different zooms were chosen based on the morphology. A red cross is shown at the phase center of each image.}
    \label{fig:jets}
\end{figure*}

Twenty-two of our BAT AGN have extended, morphologically diverse 22~GHz emission that is not jet-related. They are shown in Figure~\ref{fig:extended} and Figure~\ref{fig:jets}, respectively. Among the non-jets are examples of smooth emission extending outwards from the core (e.g., CGCG~493-002), clumpy emission (e.g., ESO~549-G049), and star formation rings (e.g., IC~2637). Although we will show that the emission in these objects is due to star formation, it is important to note that AGN are capable of driving outflows other than classical, collimated radio jets. Such outflows can interact with the environment to produce radio emission. Two of our extended objects, NGC~2992 and NGC~4388 are known to be hosts to such outflows \citep{veilleux99,veilleux01}. For consistency with the overall criteria described in this section, they remain in the ``extended" sample.

Determining whether extended radio emission is due to star formation or to a jet/AGN-driven outflow thus requires more detailed analysis than a qualitative look at morphology. If a literature search does not indicate a known, well-studied outflow or jet, we overplot the radio emission on archival optical images. If the radio emission is located preferentially outside of and perpendicular to the galaxy plane, the object would most likely be an outflow. We also check whether the observed radio emission is roughly consistent with that expected from the star formation component of the infrared \emph{Herschel} emission (see Section~\ref{model}): a significant overage is indicative of a jet. Finally, we can look carefully at the higher ($\sim0.3$\arcsec) resolution B-array images, if available, to further resolve the morphology.

\begin{figure*}
    \centering
    \includegraphics[width=\textwidth]{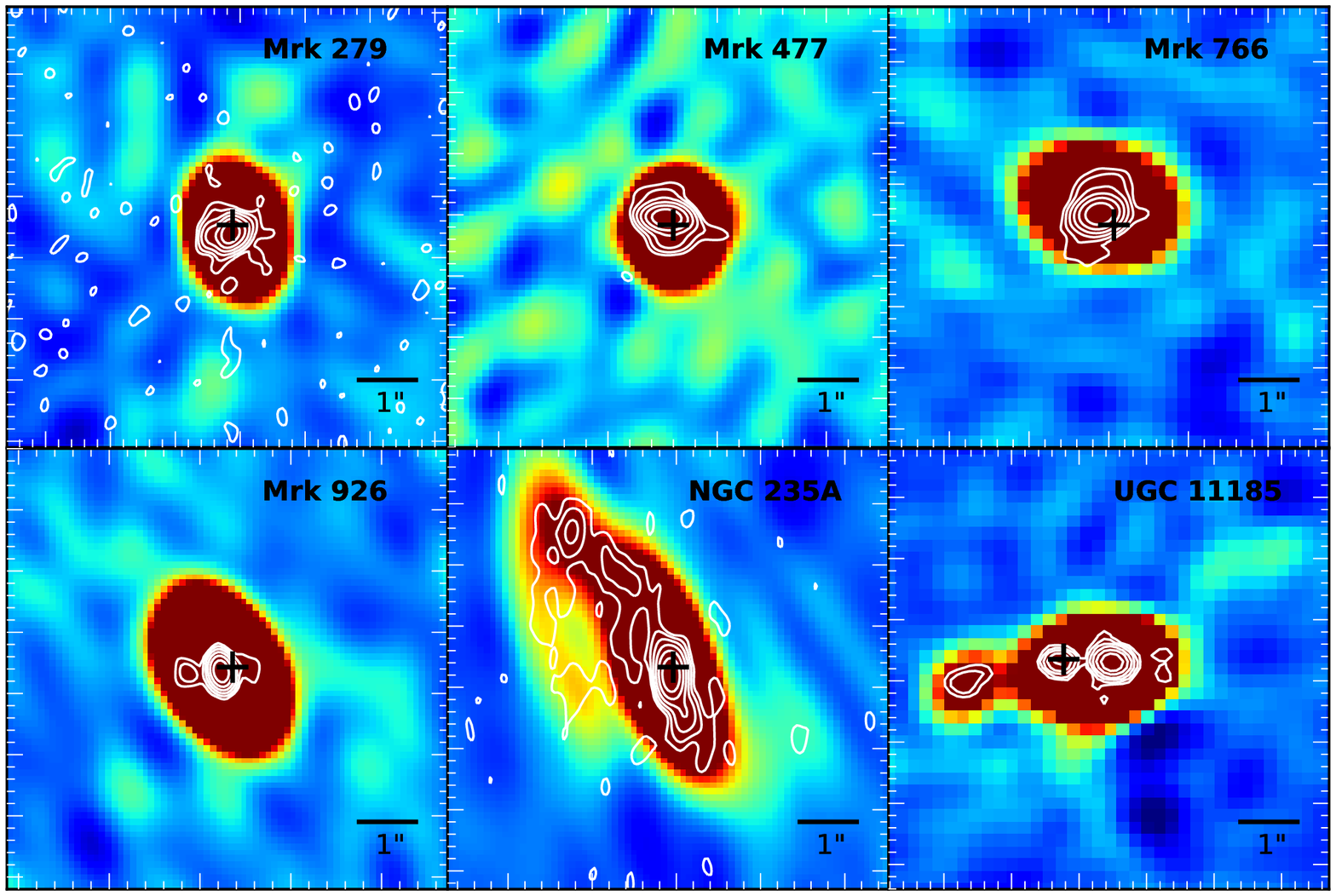}
    \caption{C-array images with 1\arcsec~resolution are shown in color in the background, while foreground contours show B-array 0.3\arcsec~resolution images overlaid. These six objects are those for which the B-array images added morphological information to that obtained from C-array. Scalebars are shown in the bottom right of each image, and a black plus sign is given at the phase center of each B-array observation.}
    \label{fig:bvc}
\end{figure*}

Only six of our B-array follow-up images contained information which might affect the morphological classification of the objects in question. A comparison of the C-array and B-array images is shown in Figure~\ref{fig:bvc}. All but four of our sources categorized as core-dominated remained compact at the B-array's higher resolution. The four exceptions have slightly extended morphology in B-array: Mrk~279, Mrk~477, Mrk 766, and Mrk~926. For consistency with the rest of the analysis in which we have defined a core-dominated source as having no extended structure in our 1\arcsec~resolution images, we do not alter their classification here. NGC~235A has a C-array morphology that could be interpreted as jet-like or star formation, while its B-array morphology shows significantly more clumping. Additionally, its extended radio flux density agrees with that predicted from the infrared star formation. These things together led us to classify it as an extended star formation object. Finally, UGC~11185 has ambiguous C-array morphology but is linear and jet-like in its B-array image, while also having an extended radio flux density in excess of the infrared star formation. We have therefore classified it as a jet.


There are two further objects that required greater scrutiny: 2MFGC~02280 and Mrk~79. Both have roughly linear structures that initially indicated a jet-like classification. However, their \emph{Herschel} PSF-subtracted images have excess infrared emission roughly co-aligned with the radio structure, and their extended flux densities are consistent with that expected from star formation. These factors together led us to classify these as extended, rather than jet-like. However, the reader should bear in mind these considerations throughout the work.

We note that all objects with a significant excess of radio emission over that predicted from the infrared star formation have linear, jet-like morphologies (see Section~\ref{condon}). The amount of extended radio emission compared to that predicted from the \emph{Herschel} observations was used to help classify only the small handful of cases where the morphology was not clearly jet-like or indicative of star formation, not as a primary classifying factor. We also note that while morphology cannot reliably differentiate between a jet and an AGN-driven wind \citep{harrison15}, we only wish to distinguish between radio emission from extended star formation and from AGN-related sources; those sources may be either jets or AGN-driven outflows without affecting our general analysis.

We conclude by noting that only six of our objects are host to arcsecond-scale jets. Importantly, this is only a small fraction of our sample; it is therefore inaccurate to assume that the unresolved emission in lower resolution surveys such as NVSS ($\theta\sim45$\arcsec) and FIRST ($\theta\sim5$\arcsec) is due entirely to scaled-down cousins of the large synchrotron jets of radio-loud AGN. See Section~\ref{origin} for a more detailed discussion of our observations' implications for the origin of radio emission in radio-quiet AGN.


\section{Testing the FIR-Radio Correlation}
\label{results}

\subsection{Radio and IR Flux Decomposition}
\label{model}

Far-infrared (FIR) continuum flux is frequently used as a star formation tracer, and it is nearly always assumed that the FIR emission comes exclusively from star formation, with no contribution from the AGN. However, our recent \emph{Herschel} observations have shown that the AGN can contribute significantly to the mid-to-far IR \emph{Herschel} SEDs. We have successfully separated the AGN and star formation components using SED decomposition for all 313 \emph{Swift}-BAT AGN with \emph{Herschel} data, which is discussed in detail in the appendix of \citet{shimizu15}. We provide a brief description here, including example SED fits for the reader.

\begin{figure*}
    \centering
    \includegraphics[width=\textwidth]{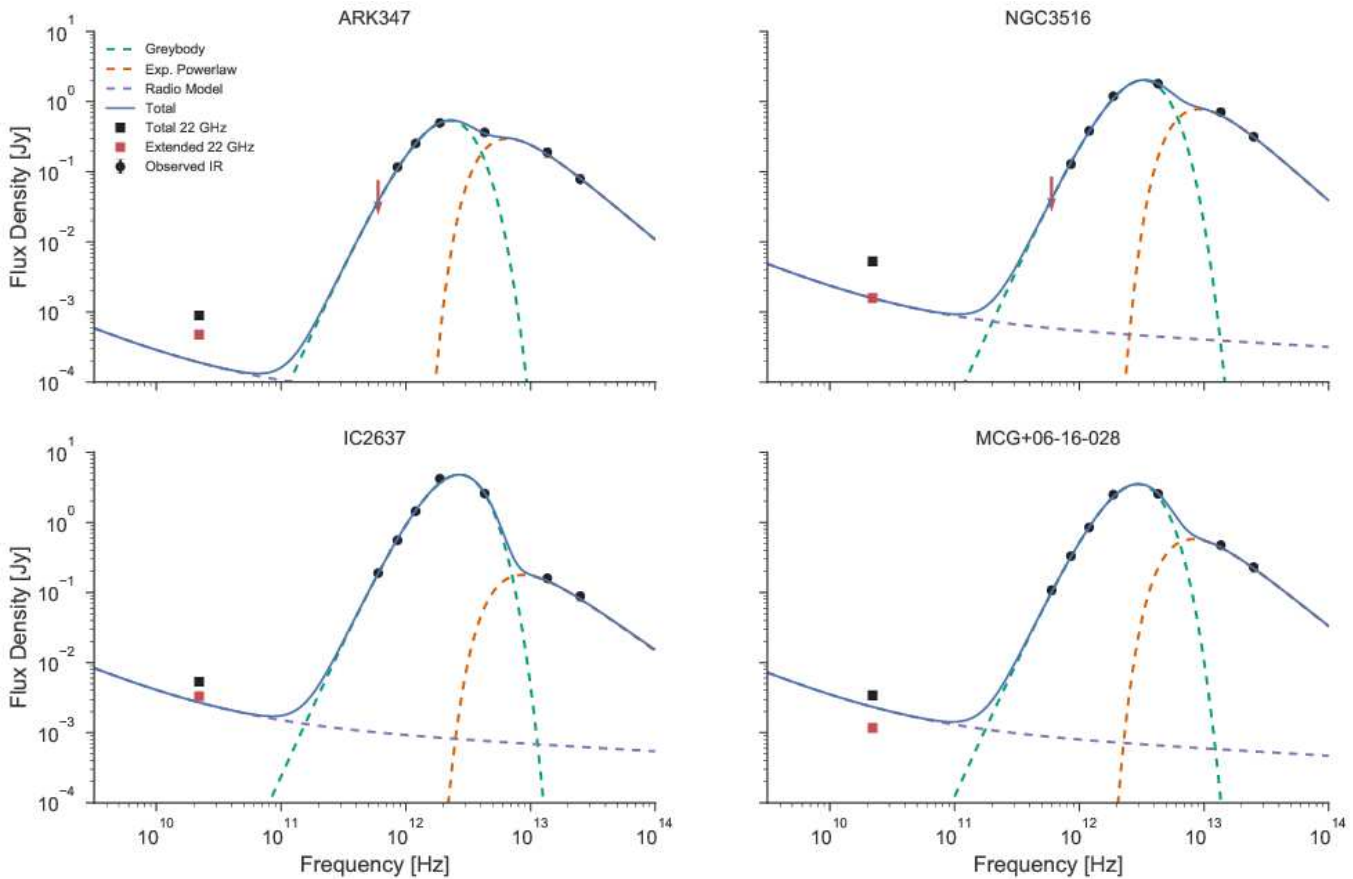}
    \caption{Decompositions of the \emph{Herschel} FIR SEDs for four representative objects in our sample: Ark~347, a core-dominated object with higher radio flux than predicted with a likely AGN contribution to the core;  NGC~3516 and IC~2637, extended cases where the model prediction matches the observed extended flux; and MCG+06-16-028, an extended object where there must be star formation within the compact core, since the extended emission alone falls below the prediction. The modeled AGN contribution is given as an orange dashed line, and the contribution from star formation is shown as a green dashed line. The purple dashed line shows the radio flux density predicted from the FIR-radio relation from the star formation component. Pink arrows denote upper limits. Error bars on the observed infrared points are smaller than the points in this log-log scaling.}
    \label{fig:samplefits}
\end{figure*}

Briefly, we are able to model the far-IR \emph{Herschel} spectra using two components, following the methodology of \citet{casey12}. The first component is an exponentially cut-off mid-IR power law to model dust in the torus heated by the AGN as the superposition of many hot dust blackbodies, and has the form $F_{\lambda}\propto\lambda^{\alpha}\rm{e}^{-(\lambda/\lambda_{turn})^{2}}$. The second component is a standard modified greybody, $F_{\nu}\propto\nu^{\beta}B_{\nu}(T)$, with dust emissivity spectral index $\beta$ and temperature $T$, which models well the FIR emission from star formation. The fitting was done using a Levenberg-Marquardt $\chi^2$-minimization algorithm and utilized all IR wavebands with at least a $5\sigma$ detection. As a sample of the fitting results, we show four of our BAT AGN SEDs in Figure~\ref{fig:samplefits}. 


To predict the 22~GHz emission due to star formation, we assume all of our sources follow the empirical FIR-radio correlation found by \citet{condon92}. Equation~14 from \citet{condon92} allows us to convert the FIR emission into a 1.4~GHz flux density. We use 60 and 100~$\mu$m flux densities based only on the best fit modified greybody component from our SED modeling. This removes any AGN contribution and provides a pure star-forming FIR flux density. In order to extrapolate from 1.4~GHz to 22~GHz, we first use the approximation of \citet{condon90}:

\begin{equation}
\frac{S}{S_T} \sim 1+10(\frac{\nu}{\mathrm{GHz}})^{\alpha_{\mathrm{brem}}-\alpha_{\mathrm{synch}}}
\end{equation}

to estimate the fraction of bremsstrahlung ($S_T$, thermal emission) and synchrotron radiation at 1.4~GHz. We then extrapolate both components to 22~GHz assuming a spectral index of $\alpha_{\mathrm{synch}}$=0.8 and $\alpha_{\mathrm{brem}}$=0.1 for the synchrotron and bremsstrahlung components, respectively, where flux density $S_{\nu}\sim \nu^{-\alpha}$. 

As a sample of the fitting results, we show four cases in Figure~\ref{fig:samplefits}: an object with observed radio emission exceeding that expected from SF (Ark~347), in which case we may assume the excess arises from the AGN; two examples in which the extended radio emission matches that predicted for pure star formation (IC~2637 and NGC~3516); and one case for which the predicted flux density falls between the total observed flux density and that of the extended emission, in which we can assume that there is additional star formation within the unresolved core, which was removed with the core subtraction (MCG+06-16-028). The next section examines how well the canonical FIR-radio relation predicts this extended emission for our entire sample.

\subsection{The FIR-Radio Correlation for BAT AGN}
\label{condon}

Our high resolution maps allow us to spatially decompose the emission into circumnuclear extended structure and an AGN nuclear component, which we can then compare to the star formation flux density expected from the canonical and oft-used FIR-radio correlation for normal galaxies from \citet{condon92}. This allows us to determine whether star formation in the immediate environment of an active nucleus differs substantially from that in the wider host.

As described in Section~\ref{sec:fluxes}, in order to fully capture the extended emission we subtract the core flux density measured in the 1\arcsec~image from the total 6\arcsec~flux density. This enables us to capture low surface-brightness emission that may not have been apparent in our higher resolution maps. It is this extended flux density, $S_{\rm{ext}}=S_{6\arcsec}-S_{\rm{core}}$, that we test against the radio flux density expected from SF via the \citet{condon92} relation, using the FIR SF component calculated with our SED modelling (see Section~{\ref{model}). The values are given in Table~\ref{t:tab1}.


\begin{figure*}
    \centering
    \includegraphics[width=\textwidth]{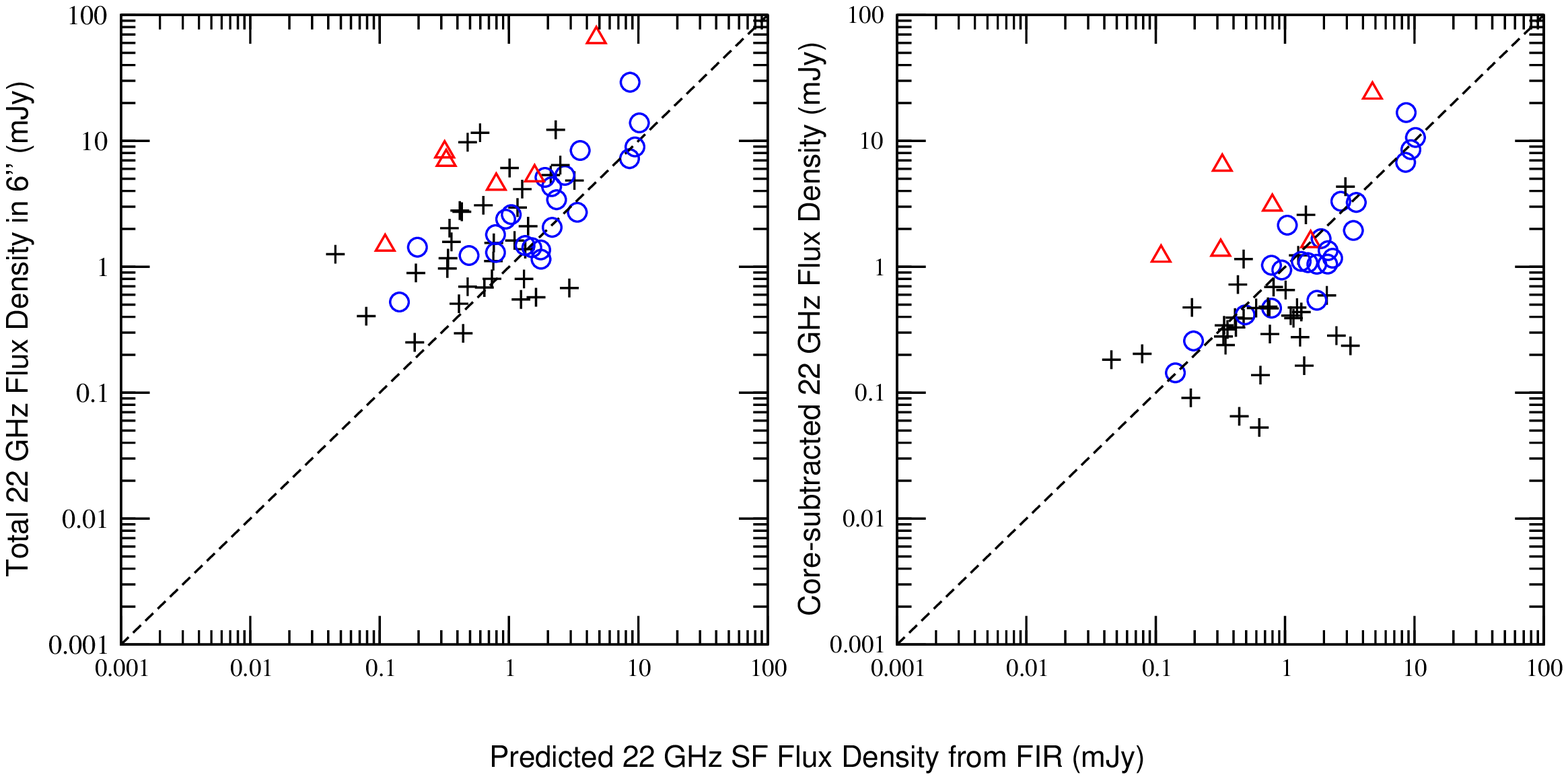}
    \caption{The observed 22~GHz flux density versus the predicted flux density from star formation based on our SED fitting (Section~\ref{model}) for the full 6\arcsec~taper (left) and for the 6\arcsec~taper with the unresolved 1\arcsec~core flux density removed (right). The dashed line is 1-to-1. Symbols correspond to jet-like sources (red triangles), extended sources (blue circles), and compact sources (black plusses).}
    \label{fig:beforeandafter}
\end{figure*}


Figure~\ref{fig:beforeandafter} plots the observed radio flux density compared to that predicted from star formation both before (left) and after (right) removal of the central core source. If we consider the entire flux density in the 6\arcsec~image, the radio flux density is systematically higher than that predicted from the star-formation component of the FIR SED. The BAT AGN therefore lie above the FIR-radio correlation when the total 22~GHz radio flux density is included. This is a similar conclusion to that reached by \citet{zakamska16}, who found that radio-quiet quasars and star-forming galaxies fall on two distinct FIR-radio relations, with the quasars having substantially higher radio fluxes. Older studies like \citet{sopp91} found that radio-quiet Seyferts tend to lie on the same FIR-radio luminosity relation as inactive late-type galaxies, albeit in much lower resolution observations. \citet{rosario13} also found that radio-quiet Seyferts occupied a similar mid-IR/radio phase space as normal star forming galaxies. However, all of these samples were optically-selected and may suffer from biases not present in our hard X-ray selected sample.  \citet{wong16} found that their sample of BAT AGN lie on the FIR-radio correlation; however, they used the total 60$\mu$m luminosity from \emph{IRAS}, rather than decomposing the full FIR SED into an AGN and SF component, in order to directly compare to the 1.4~GHz vs. 60$\mu$m luminosity relation from \citet{yun01}. Had they removed any AGN contribution to their $60\mu$m luminosities, the radio flux densities observed would have been above what was expected from the FIR-radio correlation, consistent with our result. 

The right panel of Figure~\ref{fig:beforeandafter} demonstrates the effect of removing the core radio component. The sources behave differently based on morphology. Most of the jets, as expected, remain above the expected emission from star formation. The core-dominated sources fall both above and below, but mainly below. Objects that have higher flux densities than predicted can be attributed to radio jet/outflow emission, which is perhaps not surprising since these are core-dominated sources. Those with less emission than predicted can be attributed to over-subtraction of star formation, since some of the core emission is likely due to star formation. \citet{alonso16} showed recently that a number of Seyferts harbor star formation within an unresolved $\sim$0.3\arcsec~point source in mid-IR imaging, so it is likely that many of our radio cores contain some star formation. 

Finally, one can see that the majority of objects with extended star formation morphology lie strikingly near the predicted value from the infrared star formation component. This information is also presented as a histogram in Figure~\ref{fig:histo}, for easier visualization. The blue sample in the histogram (the objects with extended morphology) cluster around a ratio of detected vs. predicted flux density of unity. We therefore conclude that once the unresolved AGN contribution to the 22~GHz and FIR flux is removed, the star formation within 75-1000~pc of the nucleus adheres to the predictions of the canonical FIR-radio relation for our sample of radio-quiet, X-ray selected AGN. Our conclusion is consistent with the results of \citet{baum93}, but inconsistent with the result of the subtraction of $\sim0.1$\arcsec~cores by \citet{roy98}, who found that the core removal did not improve the relation. However, inspection of their Figure~1 shows that while the scatter in the FIR-radio correlation did not improve after core subtraction, the sources do settle more evenly along the line, similar to our results. Therefore, although the radiation fields, star formation histories, or other factors may be different in the central few hundred parsecs than in the widespread galaxy, it does not significantly affect the physics which gives rise to the FIR-radio relation, and the infrared and radio properties of circumnuclear star formation in AGN is not manifestly different than in star forming galaxies. 

Our results show that the FIR-radio relation applies in AGN, including their inner regions. They also confirm that the star formation component can be reliably decomposed and measured from the FIR SED. However, radio emission should not be used in high-redshift radio-quiet AGN to estimate the star formation rate, since spatial resolution will be inadequate to separate emission due to star formation and the AGN.

\section{The Origin of Radio Emission in Radio-Quiet AGN}
\label{origin}

\subsection{The $L_X / L_R$ Correlation}
Once it was established that even radio-quiet AGN tend to have radio cores, the origin of this core emission became a matter of debate. One early school of thought was that all of the radio emission in radio-quiet cores came from the same sort of relativistic jets that power the large, striking plumes of synchrotron emission seen in the Fanaroff-Riley classes \citep{fanaroff74}. However, if all black holes are capable of launching synchrotron jets, and the emission from radio quiet AGN results simply from small, weak versions of such jets, one expects that radio loudness would be a smooth continuum across large AGN samples. The existence of such a smooth distribution was in doubt, since early studies concluded that the AGN radio-loudness distribution was bimodal \citep{strittmatter80, kellermann89}. More recent work, however, has shown these claims to be incorrect \citep[e.g.,][]{condon13}. Amongst luminous quasars, deep radio surveys have suggested a smooth distribution \citep{white00, cirasuolo03} and \citet{brinkmann00} saw no evidence for a bimodal luminosity distribution in the cross-matched ROSAT-FIRST catalog. Finally, \citet{ballo12} and \citet{bonchi13} found no evidence for a bimodal distribution in large samples of hard X-ray selected AGN of various Seyfert type. Although not bimodal, the radio luminosity distribution can be explained by the superposition of two populations, powered by different phenomena. \citet{kimball11} and \citet{condon13} have shown that the radio-faint population may originate from star formation in the radio-quiet QSO host, an idea also put forward by \citet{sopp91} and \citet{padovani11}. \citet{padovani15} found two very distinct populations of radio-quiet and radio-loud AGN in the \emph{Chandra Deep Field}-South VLA sample, described by differing characteristic luminosity functions and Eddington ratios. \citet{padovani15} also showed that star forming galaxies begin to dominate the faint radio sky at about 0.1~mJy, which is the same value at which radio-quiet AGN begin to outnumber radio-loud AGN. 


\begin{figure}[ht]
\begin{center}
\includegraphics[width=0.45\textwidth,trim={0.5cm 0 0 0},clip]{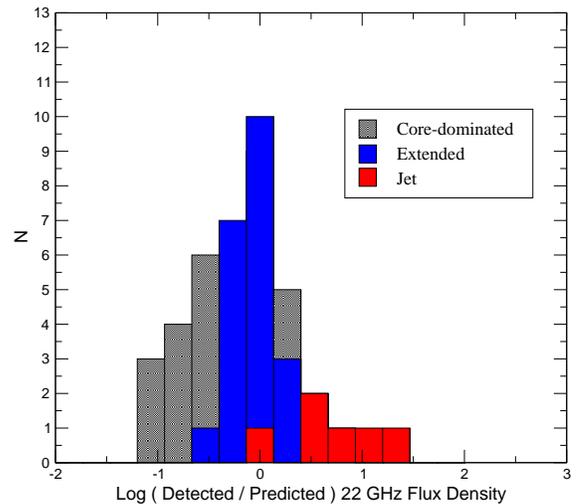}
\figcaption{Histogram showing distribution of values of the ratio of predicted SF radio flux density from SED fitting to measured radio flux density in extended emission for various source morphologies. Objects near the center point are those for which the predicted value was closest to what was measured. \\
\label{fig:histo}}
\end{center}
\end{figure}

However, \citet{zakamska16} found that star formation alone was insufficient to explain the core radio emission in their sample of radio-quiet quasars. Our data allow us to isolate the core emission, having directly measured and removed the emission from extended star formation. This enables us to study the correlation of the nuclear radio luminosity with properties such as black hole mass and accretion rate and how they might constrain the physical origin of radio-quiet emission.


\begin{figure*}
    \centering
    \includegraphics[width=\textwidth]{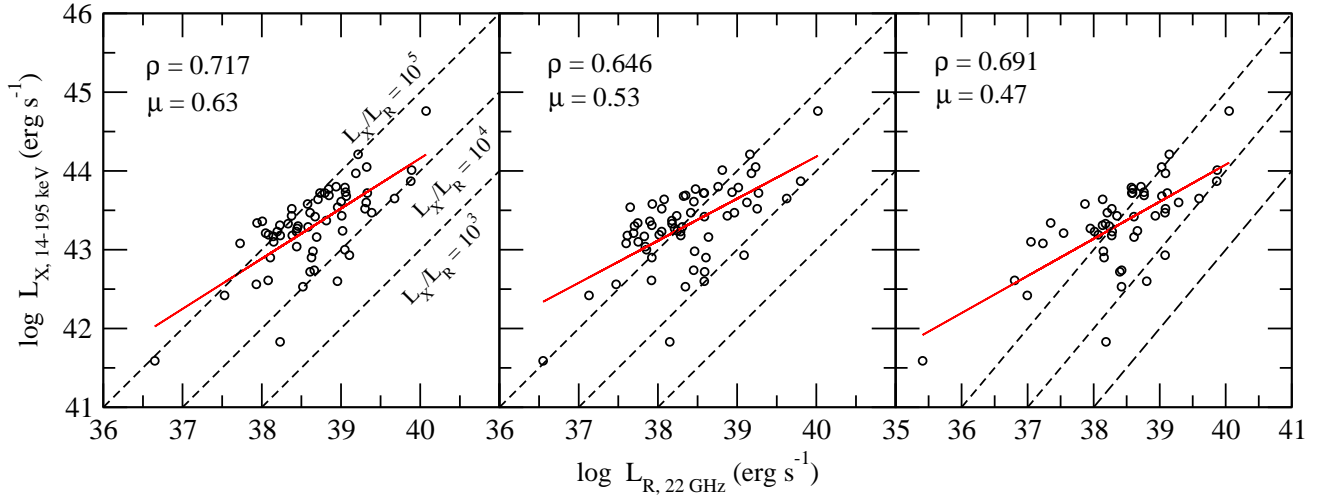}
    \caption{Hard X-ray versus 22~GHz luminosities for the total emission in 6\arcsec (left panel), the compact core emission density only (middle panel), and the 6\arcsec total emission minus that predicted from star formation from the infrared SED fits(right panel). The red lines show the best linear fit to the data. Dashed lines show the $L_X / L_R = 10^{5}$, $10^{4}$, and $10^{3}$, for comparison with previous studies (see text). The slope of each best-fit line ($\mu$) and the Spearman rank correlation coefficient ($\rho$) are shown in each panel. For our sample sizes, the 1\% probability of a chance relationship corresponds to $\rho\sim0.3$, which is comfortably exceeded in each case.}
    \label{fig:radioxray}
\end{figure*}


There are many reasons to expect a correlation between the X-ray and radio properties of an accreting compact source: the X-ray emission in radio-loud AGN could be interpreted as Compton up-scattered radio synchrotron photons; there are theoretical models postulating a fundamental connection between accretion flows and jet formation; accretion onto compact objects involves both hot accretion flows manifesting in synchrotron radiation of relativistic electrons as well as the launch of jets or outflows emitting synchrotron emission in the radio band; and so on. Indeed, an approximately linear correlation between X-ray and radio luminosity has been observed for many years, over a wide range of AGN bolometric luminosities \citep{canosa99,brinkmann00, salvato04, panessa07, ballo12}. \citet{laor08} found that the G\"{u}del-Benz relation for coronally active stars \citep[$L_{\mathrm{R}} / L_{\mathrm{X}} \sim 10^{-5}$,][]{gudel93} also held for ultra-luminous X-ray sources (ULXs), Seyfert nuclei and radio-quiet quasars, and suggested that therefore the origin of the $L_{\mathrm{R}} / L_{\mathrm{X}}$ relation is likely coronal. Further, \citet{behar15} found that the 95~GHz emission in a sample of eight radio-quiet AGN followed a $L_{\mathrm{R}} / L_{\mathrm{X}} \sim 10^{-4}$ relation, but that accounting for X-ray absorption could bring their sample into alignment with the coronal $10^{-5}$ value. Radio-loud AGN lie substantially above the relation, with $L_{\mathrm{R}} / L_{\mathrm{X}} \sim 1$, implying different physical origins \citep[e.g.,][]{behar15,balmaverde06}. 

We first explore the correlation between the radio and X-ray luminosities. The 14-195~keV luminosities are taken from the \emph{Swift}-BAT catalog. The radio luminosities are given as $\nu L_{\nu}$, for consistency with previous studies \citep[e.g.,][]{laor08}. Figure~\ref{fig:radioxray} shows the correlations between  $L_{\mathrm{X, 14-195~keV}}$ and $L_{R, 22~\mathrm{GHz}}$ for the full 6\arcsec~integrated emission (i.e., including the extended star formation; left panel), the compact core only (middle panel), and  the integrated 6\arcsec~emission minus that predicted from star formation by the infrared SED fitting (right panel). Regarding those luminosities given in the right panel, since we have shown that the infrared decomposition correctly predicts the radio emission from star formation, the difference between the total luminosity and the prediction for star formation must be attributable to the AGN. The Spearman rank correlation coefficients ($\rho$) and the slopes of the regression lines ($\mu$) are also given. We note that the correlation coefficient for all three panels is very close to that found by \citet{laor08} for radio-quiet quasars ($\rho=0.71$, see their Figure~2, top panel). The Spearman correlation coefficients all comfortably exceed the 1\% probability chance value ($\rho_{0.01} \sim0.35$). 

Having established the correlations, we can compare the luminosity ratios to those predicted by various models. Our sources cluster between the $L_{\mathrm{X}} / L_{\mathrm{R}} \sim 10^5$~and $10^4$~lines for the full 6\arcsec~emission, and around the $L_{\mathrm{X}} / L_{\mathrm{R}} \sim 10^{5}$ line for the core luminosity and the total minus the prediction for star formation. This is consistent with the coronal relationship found by \citet{laor08} and \citet{behar15}. None of the sources, even those with miniature jets (those in Figure~\ref{fig:jets}) are anywhere near $L_{\mathrm{X}} / L_{\mathrm{R}} \sim 1$, despite a significant radio component coming from the jets. This may be because these jets are not scaled-down versions of the large, collimated relativistic radio jets studied in radio-loud samples, but are instead AGN-driven outflows with radio emission from either shocks or magnetic reconnection in a wind \citep{miller06,fukumura10}. It should also be noted that the slopes of the best-fit relations are not significantly different in the panels showing the entire 6\arcsec~emission than in the core measurements. We therefore conclude that the $L_{\mathrm{X}} / L_{\mathrm{R}}$ relationship does not successfully distinguish between AGN-driven outflows (not including well-collimated relativistic jets) and star formation when interpreting radio emission in radio-quiet AGN. 

\subsection{The Fundamental Plane of Black Hole Activity}
A scattered correlation between the radio luminosity, the X-ray luminosity, and the black hole mass in accreting systems, dubbed the ``fundamental plane of black hole activity," has also been seen in many studies \citep{merloni03, falcke04, gultekin09}. It takes the following form:

\begin{equation}
\mathrm{log} ~L_{R} = \xi_{\mathrm{RX}}~\mathrm{log}~L_{X} + \xi_{\mathrm{RM}}~\mathrm{log}~M + b_{\mathrm{R}}
\end{equation}

\citet{merloni03} found values of $\xi_{\mathrm{RX}} = 0.60 \pm 0.11$ and $\xi_{\mathrm{RM}} = 0.78^{+0.11}_{-0.09}$ for the coefficients using a multivariate regression analysis, with a dispersion $\sigma_{\mathrm{R}} = 0.88$. They go on to use these measured values to constrain the physics of the accretion flow, following a scale-invariant disk-jet coupling model, and conclude that the observations are most consistent with radiatively inefficient accretion flows, such as ADAFs, as the source of the X-ray emission in objects where $L_{bol} / L_{Edd} \sim$ a few percent. This scale-invariant model assumes that all of the accretion disk physics can be reduced to a single observable quantity: the radio spectral index. \citet{merloni03} provide an excellent summary of this analysis in their Section~5, while the full treatment can be found in \citet{heinz03}. Of course, such interpretations are based on the assumption that the radio emission comes from a scaled-down jet, which is difficult to reconcile with both the fact that AGN seem to follow the G\"{u}del-Benz relation and our own discovery that only a small fraction of BAT AGN have jet morphology at high resolution. This is corroborated by studies in which the core emission in the small number of radio-quiet quasars with VLBI imaging remains unresolved even at milliarcsecond resolutions \citep[e.g.,][]{ulvestad05}; however, there are not currently any similar VLBI studies of large AGN samples in our luminosity range. \citet{merloni03} used 5~GHz measurements from a broad literature sample, while our sample is at 22~GHz, as well as using 2-10~keV X-ray luminosities as opposed to our 14-195~keV BAT luminosities, so direct comparison of the observed coefficients is not appropriate. However, we here perform a parallel analysis to theirs, and propose that our ultra-hard X-ray luminosities may provide a more pure estimate of the true X-ray power of the AGN cores than the 2-10~keV emission, and may perhaps be more appropriate for fundamental plane predictions.

The matter of measuring black hole masses in AGN is tricky: $\mathrm{M_{BH}}-\sigma_{*}$ measurements are notoriously unreliable \citep[e.g.,][]{dasyra07}, while dynamical methods are impossible for AGN at our sample's distance. In order to minimize confusion, we have decided to use only the most reliable methods of $M_{BH}$ measurement available. In order of reliability, these are: water maser emission, reverberation mapping, and the width of the broad Balmer emission lines. Two of our objects have water maser mass measurements: NGC~1194 \citep{greene14} and NGC~4388 \citep{greene10}. Nine of our AGN have reverberation estimates in the AGN Black Hole Mass Database \citep{bentz15}. Measurements based on the doppler-broadening of the Balmer emission lines are less reliable, being subject to inclination modeling effects \citep{collin06, peterson14}. However, we are including these measurements for the five of our Type~1 AGN with values in the literature to maximize sample size. The masses used for each object and the references from which we took the measurements are given in Table~\ref{t:tab2}, along with the other parameters used to calculate the predicted fluxes. We have used these masses, our 22~GHz core luminosities and our BAT X-ray luminosities in Equation~2 to calculate the predicted radio luminosity for the scale-invariant jet model of \citet{heinz03}. The result is shown in Figure~\ref{fig:fplane}. Our observed 22~GHz core luminosities lie on a relation with a slope roughly consistent to that predicted by the fundamental plane at 5~GHz, but are systematically below the predictions. If the 22~GHz luminosities are systematically fainter than at 5~GHz, it is possible that the objects would shift upwards onto the relation. However, \citet{merloni03} point out that the scatter in the fundamental plane is likely to be explained by inherent scatter in the radio spectral indices $\alpha$, so we do not attempt to scale our 22~GHz measurements to 5~GHz as we do not have spectral index information for the full sample. The fact that our objects follow a similar relation to the fundamental plane relation is consistent with the recent result at 1.4~GHz from \citet{wong16}. In short, the 22~GHz core luminosities of the small subset of our objects with black hole mass measurements are systematically below, but conform to a similar relation, to those predicted by the scale-invariant jet model of \citet{heinz03}, but are also broadly consistent with the coronal model supported by \citet{laor08}. 


\begin{figure}[ht]
\begin{center}
\includegraphics[width=0.45\textwidth,trim={1cm 0 0 0},clip]{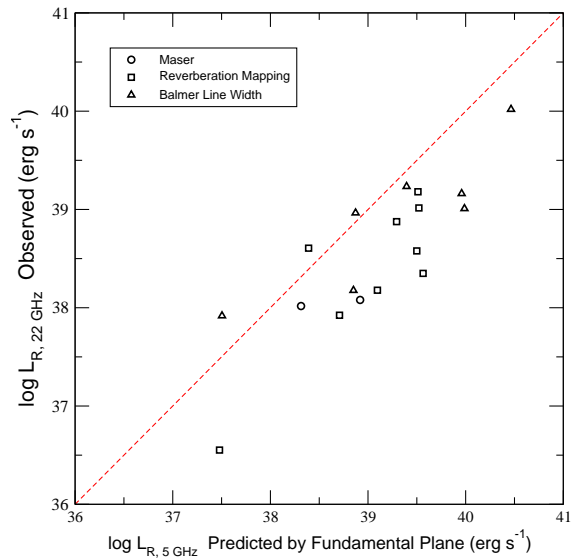}
\figcaption{Plot of the core radio luminosities observed in our sample versus those predicted by the fundamental plane of black hole activity, with symbols differentiating the black hole mass measurement method used. The red dashed line indicates a 1-to-1 correspondence.
\label{fig:fplane}}
\end{center}
\end{figure}

\section{The Main Sequence of Star Formation and Feedback}
\label{mainseq}

There is a well-studied linear relationship between the total stellar mass of normal star-forming galaxies and their star formation rates (SFRs), which holds both locally and at moderate redshifts \citep[e.g.,][]{brinchmann04, salim07,rodighiero10}, and maybe even at high redshifts \citep[e.g.,][]{heinis14, pannella15}. Since AGN host galaxies are well-known to lie in the otherwise sparsley-populated region between the blue, star forming sequence and the ``red and dead" ellipticals on color-mass plots \citep[e.g.,][]{nandra07,silverman08,schawinski09a}, it was long thought that AGN host galaxies would fall below this ``main sequence of star formation," since they should be in the process of heating and evacuating their molecular gas. Many of the first main sequence (MS) studies to include AGN did not find this, however; instead, AGN were found to lie mainly on the MS \citep{mullaney12, rosario13}. Such a finding supports the idea of co-evolution of the host galaxy and the AGN, rather than ongoing interaction. Other studies have found that AGN hosts do indeed fall below the main sequence \citep{salim07,ellison16}, and that their loci on the SFR-$M_*$ diagram are nearly perpendicular to the main sequence \citep{leslie16}. \citet{shimizu15} found the same result for the BAT AGN: that they lie systematically below the star formation main sequence, and interpret this as evidence that AGN feedback reduces the amount of cold gas available for star formation and reduces the SFR of the entire galaxy. 

\subsection{Overview of Stellar Mass and SFR Measurement}
\label{herschelsfr}

\citet{shimizu15} compared the BAT AGN to the star formation main sequence as calculated from the \emph{Herschel} Reference Survey \citep[HRS;][]{boselli10}, a complete sample of $\sim$300 galaxies with a wide variety of morphological types and with photometry matching that of the BAT sample, and the \emph{Herschel} Stripe 82 survey \citep[HerS;][]{viero14}, which includes objects with higher stellar masses more consistent with the BAT sample.

The stellar masses ($M_*$) for the galaxies were calculated using the formula

\begin{equation}
\mathrm{log}(M_*/L_i) = -0.963+1.032(g - i)
\end{equation}

from \citet{zibetti09}. This was the same equation used to calculate the stellar mass from the \emph{Herschel} Reference Survey (HRS) by \citet{cortese12}. The $g-i$ color was calculated using the \citet{koss11b} photometry. This information was available for 41 of our sources.

The SFRs were calculated by fitting the \emph{Herschel} SEDs with the \citet{casey12} model (see Section~\ref{model}). Once the star formation contribution to the IR emission is isolated, the SFR is calculated using the following relation from \citet{murphy11}:

\begin{equation}
\mathrm{SFR_{IR}} = \frac{L_{\mathrm{IR}} \mathrm{[erg~s^{-1}]}}{2.57\times10^{43}},
\end{equation}

where $L_{\mathrm{IR}}$ is the total infrared luminosity from 8-1000$\mu$m. In most of our sample, the FIR emission is unresolved by \emph{Herschel}'s 6\arcsec~imaging. In our redshift range, this scale corresponds to $\sim0.5 - 6$~kpc. Our \emph{Herschel} observations were sensitive to emission beyond this scale, and when such emission was observed it was included in the global SFR calculation. If there was significant extended emission in the host galaxies that was of insufficient surface brightness to be observed, our SFRs would be underestimated. To check whether such emission exists, we located several of our sources for which the \emph{Herschel} archive had deeper observations, and then compared the radial profile from our observation against the archival one. There was no significant difference in any of the sources we compared, indicating that we are unlikely to be missing low surface brightness extended star formation in a majority of our sources. The stellar masses and SFRs of our sample used in this analysis are given in Table~\ref{t:tab3}.

\subsection{Radio Morphologies and the Main Sequence}
\label{morphms}
We can now add our morphological information to the plot in \citet{shimizu15}, since our sample was drawn from theirs. The result is shown in Figure~\ref{fig:mainseq}. With one exception (Mrk~477), all jet-dominated and core-dominated sources are either within or well below the main sequence, while objects with nuclear star formation evident in our images reside on the main sequence. 


\begin{figure}[ht]
\begin{center}
\includegraphics[width=0.5\textwidth,trim={0.3cm 0 0 0},clip]{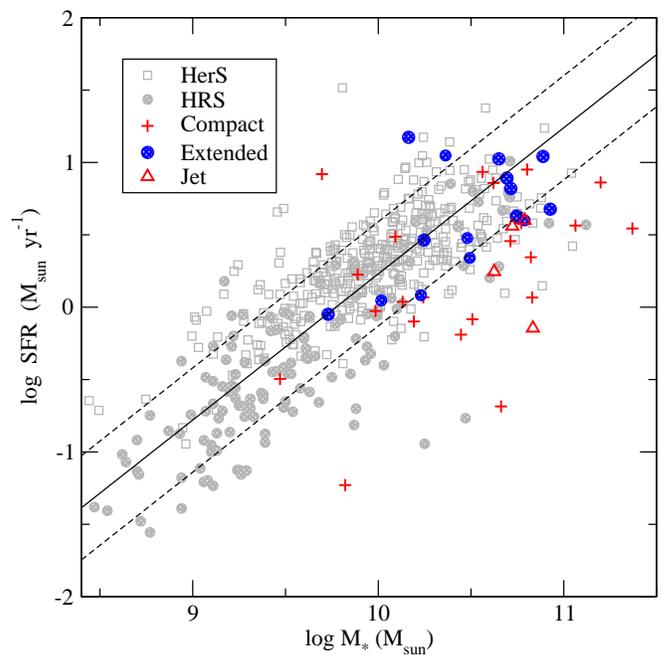}
\figcaption{The relation between stellar mass and star formation rate for the \emph{Herschel} reference survey (grey circles), \emph{Herschel} Stripe 82 survey (grey squares) and BAT AGN with compact core-dominated (red plusses), extended (blue circles), and jet-like (red triangles) 22~GHz morphology. The solid line represents the best fit to the star formation main sequence derived in S15, and the dashed lines are the 1$\sigma$ scatter.
\label{fig:mainseq}}
\end{center}
\end{figure}

Of the jet-like morphologies for which stellar masses are available, 4/4 objects fall more than $1\sigma$ below the main sequence, as well as 13/22 objects with core-dominated morphology, compared to only 3/15 objects with extended radio morphologies. All 9 sources more than $2\sigma$ below the main sequence are jet-like or core-dominated. This picture is somewhat complicated by the fact that NGC~4388 and NGC~2992, while they have extended non-jetlike morphology, are certainly dominated by AGN-driven outflows (see Section~\ref{extended}). These two objects lie within the main sequence on the plot.

Studies on the host galaxies of X-ray selected AGN, and indeed for the \emph{Swift}-BAT AGN themselves, often find that they lie in the so-called ``green valley", morphologically between the blue, star forming galaxy sequence and the ``red dead" ellipticals \citep[e.g.,][]{nandra07,schawinski09}. In \citet{shimizu15}, this conclusion was borne out: the objects lie below the main sequence of star formation, but not as far below as red ellipticals. The interpretation has often been that these galaxies are in the process of transitioning between the two regions through ongoing star formation quenching. Our results suggest that when an AGN core or a jet dominates the arcsecond-scale radio morphology, the global host star formation tends to lie below the star forming main sequence. We have proposed a large VLA follow-up program to image BAT AGN chosen explicitly to lie below the main sequence to confirm this result.

\section{Conclusions}
\label{sec:conclusion}

We have imaged 62 radio-quiet ultra-hard X-ray selected AGN using the VLA, resulting in 22~GHz maps with 1\arcsec~spatial resolution. The results and implications are described here. 

1. We find that the extended radio emission has a diverse range of morphologies, including circumnuclear star formation rings, amorphous patchy star formation, and miniature radio jets or outflows. It is therefore not appropriate to assume that all radio emission in lower-resolution imaging surveys of radio-quiet AGN is produced by small synchrotron jets.

2. After decomposing the FIR SED into AGN and SF components, we have compared the 22~GHz flux density of the spatially resolved star formation to that predicted by the FIR-radio correlation, and found that it matches very closely. We conclude that the FIR-radio correlation holds in the circumnuclear AGN environment, but that care must be taken to remove the AGN contamination \emph{from both the FIR and radio emission} before it is applied. Consequently, radio emission in radio-quiet AGN should not be used to calculate star formation rates at high redshifts, when spatial decomposition is not possible.

3. We have compared our isolated core radio luminosities to the BAT X-ray luminosities and found a strong correlation, as have previous studies. The correlation we find is consistent with a coronal origin for the radio emission in radio-quiet AGN. We have also found that the $L_{\mathrm{R}} / L_{\mathrm{X}}$ relation for our sample does not change significantly whether we are measuring the total 22~GHz radio luminosity, or only the central compact emission. The fact that $L_{\mathrm{R}} / L_{\mathrm{X}}$ does not differ wildly in our sample between jet-like morphologies (attributed to non-relativistic AGN outflows) and star formation indicates that using the ratio to distinguish between various mechanisms of radio emission in radio-quiet AGN is inappropriate, unless one is trying to identify strong relativistic jets. We have also predicted the expected radio luminosities of our sources using the fundamental plane of black hole activity, and found that our sample is consistent with the scale-invariant jet model used in constructing the fundamental plane. The jet model has numerous drawbacks, including milliarcsecond imaging failing to resolve jets as well as the consistency of $L_X/L_R$ all the way down to galactic black holes. Nevertheless, the small number of BAT AGN for which we have reliable $M_{\mathrm{BH}}$ measurements conform to the fundamental plane predictions.

4. The parent sample of our objects was found by \citet{shimizu15} to fall below the main sequence of star formation; that is, to have lower global star formation rates than normal galaxies at a given stellar mass. When we enhanced this study with our high-resolution radio morphological information, we found that the objects objects with extended radio emission tend to lie on the main sequence, while those below the main sequence exhibit compact or jet-like radio morphologies. This is the first direct connection between radio morphology and the location of AGN with respect to the main sequence of star formation, and may be evidence of ongoing star formation quenching. 

\acknowledgments

We acknowledge Peter Teuben and Kartik Sheth for assistance with CASA data reductions, as well as Heidi and Drew Medlin for their prompt and useful assistance in VLA observations scheduling. We also acknowledge Sylvain Veilleux and Ehud Behar for many helpful conversations. KLS is grateful for support from the National Radio Astronomy Observatory (NRAO). The NRAO is a facility of the National Science Foundation operated under cooperative agreement by Associated Universities, Inc. Finally, we acknowledge the very helpful comments of an anonymous referee, which have improved the manuscript.

\newpage

\begin{deluxetable}{lccccccc}
 \tablewidth{0pt}
 \tablecaption{22 GHz Properties of the BAT AGN\label{t:tab1}}
 \tablehead{
 \colhead{Name} &
 \colhead{$\alpha$} &
 \colhead{$\delta$} &
 \colhead{$z$} &
 \colhead{Predicted Flux} &
 \colhead{Core Flux} &
 \colhead{Extended Flux} &
 \colhead{Morphology} \\
 \colhead{ } &
 \colhead{ } &
 \colhead{ } &
 \colhead{ } &
 \colhead{Density } &
 \colhead{Density } &
 \colhead{Density } &
 \colhead{ } \\
  \colhead{(1)} &
 \colhead{(2)} &
 \colhead{(3)} &
 \colhead{(4)} &
 \colhead{ (mJy) (5) } &
 \colhead{ (mJy) (6) } &
 \colhead{ (mJy) (7) } &
 \colhead{(8)} \\
  }
 \startdata 
2MASX J0353+3714       	&	58.427	&	37.235	&	0.018	&	0.74	&	0.32	&	0.48	&	compact	\\
2MASX J0423+0408       	&	65.9199	&	4.1338	&	0.045	&	0.33	&	0.58	&	6.41	&	jet	\\
2MASX J0444+2813       	&	71.0376	&	28.2168	&	0.011	&	1.26	&	2.91	&	1.23	&	compact	\\
2MASX J1200+0648       	&	180.2413	&	6.8064	&	0.036	&	0.49	&	0.82	&	0.42	&	extended	\\
2MASX J1546+6929       	&	236.6014	&	69.4861	&	0.038	&	0.11	&	0.27	&	1.21	&	jet	\\
2MASX J1937-0613       	&	294.388	&	-6.218	&	0.010	&	3.55	&	5.16	&	3.24	&	extended	\\
2MASX J2010+4800	&	302.5725	&	48.0059	&	0.026	&	0.19	&	0.16	&	0.09	&	compact	\\
2MFGC 02280                   	&	42.6775	&	54.7049	&	0.015	&	1.50	&	0.34	&	1.08	&	extended	\\
ARK 347                       	&	181.1237	&	20.3162	&	0.022	&	0.19	&	0.42	&	0.48	&	compact	\\
CGCG 122-055                  	&	145.52	&	23.6853	&	0.021	&	0.35	&	1.79	&	0.24	&	compact	\\
CGCG 229-015                  	&	286.3581	&	42.461	&	0.028	&	0.08	&	0.20	&	0.20	&	compact	\\
CGCG 420-015                  	&	73.3573	&	4.0616	&	0.029	&	0.34	&	0.83	&	0.34	&	compact	\\
CGCG 493-002                  	&	324.639	&	32.085	&	0.025	&	0.20	&	1.18	&	0.26	&	extended	\\
ESO 548-G081                  	&	55.5155	&	-21.2444	&	0.014	&	1.04	&	0.46	&	2.14	&	extended	\\
ESO 549- G 049                	&	60.607	&	-18.048	&	0.026	&	3.37	&	0.77	&	1.94	&	extended	\\
IC 0486                       	&	120.0874	&	26.6135	&	0.027	&	0.79	&	0.77	&	1.03	&	extended	\\
IC 2461                       	&	139.992	&	37.191	&	0.008	&	1.62	&	0.46	&	0.69	&	compact	\\
IC 2637                       	&	168.457	&	9.586	&	0.029	&	2.70	&	2.01	&	3.32	&	extended	\\
IGR J23308              	&	352.696	&	71.336	&	0.037	&	0.41	&	0.11	&	0.40	&	compact	\\
IRAS 05589             	&	90.5446	&	28.4728	&	0.033	&	0.42	&	2.46	&	0.33	&	compact	\\
MCG -01-30-041                	&	178.159	&	-5.207	&	0.019	&	1.33	&	0.37	&	1.10	&	extended	\\
MCG +02-57-002                	&	335.938	&	11.836	&	0.029	&	0.14	&	0.38	&	0.14	&	extended	\\
MCG +04-48-002                	&	307.1461	&	25.7333	&	0.014	&	9.40	&	0.44	&	8.53	&	extended	\\
MCG +06-16-028                	&	108.5161	&	35.2793	&	0.016	&	2.33	&	2.24	&	1.17	&	extended	\\
Mrk 18                        	&	135.493	&	60.152	&	0.011	&	1.90	&	3.45	&	1.68	&	extended	\\
Mrk 198                       	&	182.3088	&	47.0583	&	0.024	&	0.79	&	0.83	&	0.47	&	extended	\\
Mrk 279                       	&	208.2644	&	69.3082	&	0.030	&	0.63	&	3.03	&	0.05	&	compact	\\
Mrk 359                       	&	21.8856	&	19.1788	&	0.017	&	1.31	&	0.53	&	0.28	&	compact	\\
Mrk 477                       	&	220.1587	&	53.5044	&	0.038	&	1.01	&	5.45	&	0.65	&	compact	\\
Mrk 590                       	&	33.64	&	-0.767	&	0.026	&	0.43	&	2.02	&	0.72	&	compact	\\
Mrk 739E                      	&	174.122	&	21.596	&	0.030	&	1.76	&	0.31	&	1.05	&	extended	\\
Mrk 766                       	&	184.6105	&	29.8129	&	0.013	&	3.20	&	4.60	&	0.24	&	compact	\\
Mrk 79                        	&	115.6367	&	49.8097	&	0.022	&	0.94	&	1.45	&	0.94	&	extended	\\
Mrk 817                       	&	219.092	&	58.7943	&	0.031	&	1.40	&	1.94	&	0.16	&	compact	\\
Mrk 885                       	&	247.451	&	67.3783	&	0.025	&	0.44	&	0.23	&	0.07	&	compact	\\
Mrk 926                       	&	346.1811	&	-8.6857	&	0.047	&	0.48	&	8.62	&	1.15	&	compact	\\
Mrk 975                       	&	18.4626	&	13.2717	&	0.050	&	0.76	&	1.26	&	0.29	&	compact	\\
NGC 1106                      	&	42.6688	&	41.6715	&	0.014	&	0.60	&	11.15	&	0.47	&	compact	\\
NGC 1125                      	&	42.918	&	-16.651	&	0.011	&	2.49	&	6.14	&	0.28	&	compact	\\
NGC 1194                      	&	45.9546	&	-1.1037	&	0.014	&	0.05	&	1.08	&	0.18	&	compact	\\
NGC 2110                      	&	88.0474	&	-7.4562	&	0.008	&	4.73	&	42.17	&	23.93	&	jet	\\
NGC 235A                      	&	10.72	&	-23.541	&	0.022	&	2.13	&	3.28	&	1.05	&	extended	\\
NGC 2655                      	&	133.9072	&	78.2231	&	0.005	&	2.29	&	12.51	&	2.59	&	compact	\\
NGC 2992                      	&	146.4252	&	-14.3264	&	0.008	&	8.63	&	12.49	&	16.78	&	extended	\\
NGC 3081                      	&	149.8731	&	-22.8263	&	0.008	&	1.10	&	1.21	&	0.41	&	compact	\\
NGC 3431                      	&	162.8127	&	-17.008	&	0.018	&	0.33	&	0.69	&	0.28	&	compact	\\
NGC 3516                      	&	166.6979	&	72.5686	&	0.009	&	1.58	&	3.70	&	1.58	&	jet	\\
NGC 3786                      	&	174.927	&	31.909	&	0.009	&	2.15	&	0.72	&	1.34	&	extended	\\
NGC 4388                      	&	186.4448	&	12.6621	&	0.008	&	10.19	&	3.26	&	10.64	&	extended	\\
NGC 513                       	&	21.1119	&	33.7995	&	0.020	&	2.92	&	0.87	&	4.33	&	compact	\\
NGC 5231                      	&	203.951	&	2.999	&	0.022	&	0.76	&	0.64	&	0.47	&	compact	\\
NGC 5273                      	&	205.5347	&	35.6542	&	0.004	&	0.64	&	0.55	&	0.14	&	compact	\\
NGC 5548                      	&	214.4981	&	25.1368	&	0.017	&	0.80	&	1.44	&	3.08	&	jet	\\
NGC 6552                      	&	270.0304	&	66.6151	&	0.026	&	2.10	&	4.76	&	0.59	&	compact	\\
NGC 7679                      	&	352.1944	&	3.5114	&	0.017	&	8.55	&	0.46	&	6.76	&	extended	\\
UGC 03478                     	&	98.1965	&	63.6737	&	0.013	&	1.34	&	0.97	&	0.44	&	compact	\\
UGC 03601                     	&	103.9564	&	40.0002	&	0.017	&	0.36	&	1.26	&	0.32	&	compact	\\
UGC 07064                     	&	181.1806	&	31.1773	&	0.025	&	1.77	&	0.61	&	0.54	&	extended	\\
UGC 08327            	&	198.822	&	44.4071	&	0.037	&	1.16	&	2.57	&	0.39	&	compact	\\
UGC 11185          	&	274.0487	&	42.6608	&	0.041	&	0.32	&	6.82	&	1.35	&	jet	\\
UGC 12741                     	&	355.4811	&	30.5818	&	0.017	&	0.48	&	0.31	&	0.39	&	compact	\\
UGC12282	&	344.7312	&	40.9315	&	0.017	&	1.24	&	0.44	&	0.48	&	compact	\\

 \enddata
 
 \tablecomments{Properties of the 22 GHz observations of our sample of BAT AGN. Columns are (1) object name, (2) right ascension, (3) declination, (4) redshift, (5) the 22~GHz flux density predicted from the star formation component of the \emph{Herschel} SED using the FIR-radio relation (see Section~\ref{model}), (6) the observed 22~GHz flux density in the compact core, (7) the observed 22~GHz flux density in full 6\arcsec~resolution image minus the core component, thereby encompassing all the extended emission, and (8) the morphological classification based on the factors described in Section \ref{morph}.}

 \end{deluxetable}

\newpage


\begin{deluxetable}{lcccc}
 \tablewidth{0pt}
 \tablecaption{Fundamental Plane Parameters\label{t:tab2}}
 \tablehead{
 \colhead{Name} &
 \colhead{Log $M_{BH}$} &
 \colhead{ $M_{BH}$ source} &
 \colhead{Log L$_{14-195\mathrm{keV}}$} &
 \colhead{Log Predicted L$_{5\mathrm{GHz}}$} \\
 \colhead{ } &
 \colhead{ } &
 \colhead{ } &
 \colhead{(erg s$^{-1}$)} &
 \colhead{(erg s$^{-1}$)} }
 \startdata 

CGCG 229-015                  	&	6.91	&	RM$^1$	&	43.31	&	38.10	\\
IC 2637                       	&	7.00	&	H$\alpha^2$	&	43.47	&	38.26	\\
Mrk 279                       	&	7.44	&	RM$^3$	&	43.97	&	38.90\\
Mrk 590                       	&	7.57	&	RM$^3$	&	43.43	&	38.68	\\
Mrk 739E                      	&	7.05	&	H$\beta^4$	&	43.37	&	38.24	\\
Mrk 766                       	&	6.82	&	RM$^5$	&	42.9	&	37.78	\\
Mrk 79                        	&	7.61	&	RM$^3$	&	43.72	&	38.89	\\
Mrk 817                       	&	7.59	&	RM$^6$	&	43.79	&	38.91	\\
Mrk 926                       	&	8.05	&	H$\beta^7$	&	44.76	&	39.85	\\
Mrk 975                       	&	7.23	&	H$\beta^8$	&	44.05	&	38.78	\\
NGC 1194                      	&	6.50	&	maser$^9$	&	43.19	&	37.70	\\
NGC 3516                      	&	7.40	&	RM$^{10}$	&	43.33	&	38.48	\\
NGC 4388                      	&	6.93	&	maser$^{11}$	&	43.64	&	38.31	\\
NGC 5273                      	&	6.66	&	RM$^{12}$	&	41.59	&	36.87	\\
NGC 5548                      	&	7.72	&	RM$^3$	&	43.69	&	38.95	\\
UGC 03478                     	&	5.90	&	H$\beta^8$	&	42.61	&	36.89	\\

 \enddata

 \tablecomments{Quantities necessary for and predicted by the fundamental plane of black hole activity as described by \citet{merloni03}. Sources of the black hole mass measurements are: $^1$\citet{barth11}, $^2$\citet{greene05}, $^3$\citet{peterson04}, $^4$\citet{koss11a}, $^5$\citet{bentz10}, $^6$\citet{peterson98}, $^7$\citet{koll10}, $^8$\citet{botte04}, $^9$\citet{greene14}, $^{10}$\citet{denney10}, $^{11}$\citet{greene10}, $^{12}$\citet{bentz14}}

 \end{deluxetable}
\newpage


\begin{deluxetable}{lcc}
 \tablewidth{0pt}
 \tablecaption{Measured Star Formation Properties\label{t:tab3}}
 \tablehead{
 \colhead{Name} &
 \colhead{Log $M_*$} &
 \colhead{ Log SFR} \\
 \colhead{ } &
 \colhead{$M_\odot$} &
 \colhead{$M_\odot$ yr$^{-1}$} }
 \startdata 

2MASX J0353+3714	&	9.89	&	0.225	\\
2MASX J0444+2813	&	10.83	&	0.067	\\
2MASX J1200+0648	&	10.93	&	0.677	\\
2MFGC 02280	&	10.49	&	0.340	\\
ARK 347	&	10.51	&	-0.083	\\
CGCG 122-055	&	10.13	&	0.037	\\
CGCG 420-015	&	10.82	&	0.345	\\
IC 0486	&	10.75	&	0.629	\\
IC 2461	&	9.47	&	-0.496	\\
MCG +04-48-002	&	10.36	&	1.048	\\
Mrk 1210	&	9.99	&	-0.026	\\
Mrk 18	&	9.73	&	-0.049	\\
Mrk 198	&	10.25	&	0.463	\\
Mrk 279	&	10.77	&	0.581	\\
Mrk 477	&	9.70	&	0.919	\\
Mrk 590	&	11.06	&	0.564	\\
Mrk 739E	&	10.65	&	1.024	\\
Mrk 766	&	10.09	&	0.488	\\
Mrk 79	&	10.79	&	0.600	\\
Mrk 817	&	10.56	&	0.935	\\
Mrk 926	&	11.20	&	0.862	\\
NGC 1194	&	10.66	&	-0.686	\\
NGC 2110	&	11.07	&	0.292	\\
NGC 235A	&	10.71	&	0.820	\\
NGC 2992	&	10.48	&	0.478	\\
NGC 3081	&	10.45	&	-0.190	\\
NGC 3516	&	10.83	&	-0.145	\\
NGC 3786	&	10.23	&	0.083	\\
NGC 4388	&	10.02	&	0.046	\\
NGC 513	&	10.62	&	0.858	\\
NGC 5231	&	10.71	&	0.457	\\
NGC 5273	&	9.82	&	-1.229	\\
NGC 5548	&	10.62	&	0.245	\\
NGC 7679	&	10.16	&	1.174	\\
NGC 985	&	10.89	&	1.042	\\
UGC 03601	&	10.19	&	-0.099	\\
UGC 07064	&	10.69	&	0.892	\\
UGC 08327	&	10.80	&	0.951	\\
UGC 11185	&	10.72	&	0.559	\\
UGC 12282	&	11.37	&	0.544	\\
UGC 12741	&	10.24	&	0.068	\\

 \enddata

 \tablecomments{Total stellar masses and star formation rates measured as described in Section~\ref{herschelsfr}.}

 \end{deluxetable}

\end{document}